\documentclass[11pt]{article} \usepackage{latexsym}
\usepackage{a4}
\usepackage[dvips]{graphics}
 \newcommand{\tr}{\mbox{$\mbox{tr}$}}
\newcommand{\Tr}{\mbox{$\mbox{Tr}$}}
\newcommand{\unit}{\textsl{\mbox{\textbf{1}}}}
\newcommand{\neu}[1]{\mbox{$#1^{\rm (n)}$}} \newcommand{\ch}[1]{\mbox{$#1^{\rm
      (ch)}$}} 
\newcommand{\dalembert}{\mbox{$\displaystyle\Box$}}

\begin{document} 
\begin{titlepage}
\mbox{ UNITU-THEP-20/1997}
  \vspace{1.5cm}


\begin{centering}
  \Large\bfseries
  Ward Identities for Yang-Mills Theory \\
  in Abelian Gauges: \\
  Abelian Dominance at High Energies \\
\end{centering}
\vspace{1cm}


\renewcommand{\thefootnote}{\fnsymbol{footnote}} \centerline{M.\ 
  Quandt\footnote[1]{Supported by "Graduiertenkolleg: Hadronen und Kerne"},
  H.\ Reinhardt\footnote[2]{Supported in part by DFG under contract Re
    856/1-3}} \renewcommand{\thefootnote}{arabic{footnote}} \vspace{1.5cm}

\centerline{ Institut f\"ur Theoretische Physik, Universit\"at T\"ubingen }
\centerline{D--72076 T\"ubingen, Germany.}  \vspace{1.5cm}


\begin{abstract}
  We consider Yang-Mills theory in a general class of Abelian gauges.
  Exploiting the residual Abelian symmetry on a quantum level, we derive a set
  of Ward identities in functional form, valid to all orders in perturbation
  theory. As a consequence, the coupling constant is only renormalised through
  the Abelian two-point function. This implies that asymptotic freedom in all
  Abelian gauges can be understood from an effective Abelian theory alone,
  which can be interpreted as Abelian dominance in the high energy regime.
\end{abstract}


\end{titlepage}


\section{ Introduction } 
\label{sec:1}

In recent years, the possibility of realising confinement through a dual 
Meissner effect, as originally proposed by {}'t Hooft and Mandelstam \cite{R2}, has
triggered a vast variety of both analytical and numerical investigations. The
magnetic monopoles necessary for the dual Meissner effect arise in the so-called
Abelian gauges proposed by {}'t Hooft \cite{R3}. Recent lattice calculations
\cite{R4} performed in these gauges have accumulated evidence in favour of
Abelian dominance and the realisation of confinement through a dual Meissner
effect.  To be more precise, the lattice calculations performed in the maximal
Abelian gauge show that about 95\% of the full string tension are produced by
Abelian gauge field configurations alone, of which about 90\% are produced
by magnetic monopoles \cite{R5}. It is important to note that, in these
lattice calculations, the sampling of the gauge field configurations is still
done with the full Yang-Mills action, so that the non-Abelian field
configurations are hidden in the weight with which the Abelian configurations
contribute to, say, the partition function.

The Abelian, and in particular, monopole dominance is interpreted as
supporting the dual superconductor picture of confinement. Furthermore, the
Abelian dominance shows that confinement properties such as the string
tension can be entirely obtained from an effective Abelian theory.

While the Abelian dominance in the string tension, i.~e.~in the low energy
sector, has been intensively studied, little is known about possible Abelian
dominance in the high-energy regime, in particular whether asymptotic freedom
can be obtained from an effective Abelian theory. In the present paper we
answer this question in the affirmative. We consider Yang-Mills theory in an
arbitrary Abelian gauge. For the residual Abelian gauge symmetry we
derive a set of Ward identities valid to all orders in perturbation theory. As
a consequence of these Ward identities, the coupling constant is renormalised
only through the Abelian two-point function, implying that in all Abelian
gauges asymptotic freedom can be understood from an effective Abelian theory
alone. This fact can be interpreted as Abelian dominance in the high energy
regime. At one loop level and in the special case of the maximal Abelian gauge
this fact was already observed in ref.~\cite{R1} and subsequently confirmed in
ref.~\cite{R6}.


\section{The Ward Identity in Abelian Gauges} 
\label{sec:2}

We consider pure $SU(N)$ Yang-Mills theory on an Euclidean four-manifold
$\mathcal{M}$ with the action given by
\begin{equation}
S_{\rm YM}[A] = \frac{1}{4}\int\limits_{\cal M}\,d^4 x\,
F^a_{\mu\nu}(x) F_a^{\mu\nu}(x).
\label{1}
\end{equation}
Here, $F_{\mu\nu} = F_{\mu\nu}^a T^a= \partial_\mu A_\nu - \partial_\nu A_\mu
+ g [A_\mu,A_\nu]$ denotes the field strength tensor of the gauge field $A_\mu
\equiv A_\mu^a T^a$, $g$ is the coupling strength and the generators $T^a$ of
the Lie algebra $[T^a,T^b] = f^{a b c} T^c$ are taken to be antihermitian and
normalised according to $\tr\left(T^a T^b\right) = -\frac{1}{2} \delta^{a b}$.

The basic idea of Abelian gauges is to remove as many non-Abelian degrees of
freedom as possible by partially fixing the gauge, leaving a theory with a
residual Abelian gauge symmetry. This is accomplished by the so-called Cartan
decomposition, $G = H \times G/H$, where $G=SU(N)$ is the gauge symmetry group
and $H\simeq U(1)^{N-1}$ denotes the maximal Abelian subgroup, spanned by a
maximal set of commuting generators $\{T^{a_0}\}$. As a convention, we will
use colour indices $a_0,b_0$ etc.~to denote the generators of this Cartan
subalgebra, while letters with a bar, $\bar{a}, \bar{b}$ etc.~are reserved for
the remaining generators.\footnote{For $G=SU(N)$ in the fundamental
  representation, we can always adjust the Cartan decomposition in such a way
  that the Abelian generators $T^{a_0}$ are diagonal and the remaining
  $T^{\bar{a}}$ have vanishing diagonal elements.}  The gauge field is then
decomposed into its diagonal and off-diagonal parts,
\begin{equation}
 A_\mu(x) = A_\mu^{b_0} T^{b_0} + A_\mu^{\bar{b}} T^{\bar{b}} \equiv
\neu{A_\mu} + \ch{A_\mu}.
\label{2}
\end{equation}
The superscripts $\ch{}$ and $\neu{}$ for "charged" and "neutral" refer to the
transformation properties under the residual Abelian gauge group $\omega =
e^{-g \theta^{b_0} T^{b_0}} \in H$:
\begin{equation}
\neu{A_\mu} \rightarrow \neu{A_\mu} + \frac{1}{g}\,\omega\cdot\partial_\mu 
\omega^{\dagger} = \neu{A_\mu} + \partial_\mu \theta \quad ; \quad
\ch{A_\mu} \rightarrow \omega\cdot\ch{A_\mu}\cdot\omega^{\dagger}.
\label{3}
\end{equation}
Under this residual symmetry, $\ch{A_\mu}$ transforms as a charged matter
field in the adjoint representation, while the diagonal part $\neu{A_\mu}$
behaves like a photon.

To fix the coset we need $N(N-1)$ conditions which leave the residual Abelian
symmetry (\ref{3}) unbroken. More specifically, we will consider charged gauge
fixing conditions of the form
\begin{equation}
\ch{\chi}[A] \equiv \chi^{\bar{a}}[A] T^{\bar{a}} = 0 \quad ; \quad
\ch{\chi}[A^\omega] = \omega\cdot\ch{\chi}[A]\cdot\omega^{\dagger},
\label{4}
\end{equation}
which are obviously invariant under the residual symmetry (\ref{3}).  This
form includes all the commonly used Abelian gauges, in particular the
"diagonalisation gauges" and the so-called \emph{maximal Abelian gauge}.

The main motivation for the use of Abelian gauges is that, besides the
appearance of magnetic monopoles, it should facilitate integrating out the
charged gauge field components $\ch{A_\mu}$ leaving an effective Abelian
theory. The quantisation of the latter still requires a gauge fixing for the
neutral photon, which in the present paper will be done by the usual Lorentz
condition
\begin{equation}
\chi^{b_0} \equiv \partial_\mu A_\mu^{b_0} = 0.
\label{5}
\end{equation}
The complete gauge fixing can be regarded as a two-step process with the
charged and neutral gauge conditions (\ref{4}) and (\ref{5}) being essentially
\emph{independent} of each other. This is justified by the standard
Faddeev-Popov (or BRST) quantisation of the combined gauge (\ref{4},\ref{5}).
Relaxing the gauge conditions in the usual way by introducing gauge-fixing
terms into the action, we obtain the generating functional in the form
\begin{eqnarray}
Z[\neu{j},\ch{j}] &=& \int {\cal D}(\ch{A},\neu{A})\, \exp\Bigg\{
- \frac{1}{\hbar} S_q[\neu{A},\ch{A}] - \frac{1}{2\hbar \alpha}
\int d^4x\,\left(\partial_\mu A_\mu^{b_0}\right)^2 + 
\nonumber \\
{} && \qquad + \int d^4x\,\, j_\mu^{b_0} A_\mu^{b_0} + \int d^4x\,\, 
j_\mu^{\bar{b}} A_\mu^{\bar{b}} \Bigg\}   
\nonumber \\
S_q &=& S_{\rm YM} + \frac{1}{2\bar{\alpha}} \int d^4x\,
\chi^{\bar{a}}\chi^{\bar{a}} - \hbar\,\Tr\ln{\sf M} .
\label{6}
\end{eqnarray}
Here, $\alpha$ and $\bar{\alpha}$ are the gauge fixing parameters in the
neutral and charged sector, respectively, and ${\sf M}$ denotes the
Faddeev-Popov matrix. In the standard gauges, the non-Abelian nature of the
Yang-Mills theory complicates the derivation of Ward identities from the
functional integral. The problems are actually two-fold:

Firstly, the variation of the gf.~term under an infinitesimal gauge rotation
involves the FP-matrix, leading to highly non-linear expressions in the
resulting identities. Secondly, the FP-determinant calculated from
\begin{equation}
{\sf M}^{ab}(x,y) = \left.\frac{\delta \chi^a[A^\Omega](x)}{\delta \varphi^b(y)}
\right|_{\varphi = 0} \equiv {\sf M}[A] \qquad ; \qquad
\Omega = \exp(- g\,\varphi^a T^a)
\label{6a}
\end{equation}
coincides with the gauge invariant Faddeev-Popov measure factor $\Delta$ only
on the gf.~mani\-fold. This means that for \emph{arbitrary} configurations $A$,
the FP determinant $\det {\sf M}[A]$ will not be manifestly gauge 
invariant \cite{R7}. To see this more explicitly, we could follow the standard 
procedure and relax the gauge condition to $\chi^a[A] = c^a$. This will not 
alter the functional form of ${\sf M}$, but the gf.~manifold and the measure 
factor $\Delta_c$ will clearly change. The gauge fixing delta function can now 
be removed by averaging over $c^a$ with a Gaussian weight, assuming that the
$c$-dependence of $\Delta_c$ is eliminated by resolving the gf.~constraint,
$c^a = \chi^a[A]$.  As a result, we find that $\det{\sf M}[A]$ is equivalent 
(for \emph{all} configurations $A$) to $\Delta_{\chi[A]}$, and the 
\emph{implicit} dependence on $A$ destroys the gauge invariance \cite{R7}.

Due to these problems in the non-Abelian case, one cannot derive simple Ward
identities directly based on the underlying gauge symmetry. Instead, one 
usually exploits the BRST invariance of the FP action leading to 
Slavnov-Taylor identities which are considerably more involved than their
counterparts in QED. For the residual Cartan symmetry, however, we can
return to simple Abelian relations even for the completely gauge
fixed YM theory. \emph{This possibility is the key property of 
Abelian gauges}.    
In fact, we will show that the quantum action $S_q$ of (\ref{6}), 
including the (coset) gauge fixing term and the FP determinant, is 
invariant under the residual symmetry (\ref{3}). From this
point of view, YM theory in Abelian gauges is exactly equivalent to an  
Abelian system of photons and charged matter, although with a non-standard
effective action.  

Let us have a closer look at the behaviour of the quantum action (\ref{6}) 
w.r.t.~the Abelian transformation (\ref{3}). The invariance of the 
coset gauge fixing term is a simple consequence of the transformation 
rule (\ref{4}). We could even take this invariance as a \emph{definition} 
for an Abelian gauge. As for the last term in  $S_q$, i.e.~the
FP determinant, we have seen above that its invariance under the gauge  
rotation (\ref{3}) is not obvious and, in fact, relies on the special 
property (\ref{4}) of Abelian gauges. This can be seen as follows:

Let $\omega = e^{- g \theta^{b_0} T^{b_0}}$ be an arbitrary Abelian gauge
transformation. In components, the gauge fixing constraints transform as
\begin{eqnarray}
\chi^{a_0}[A^\omega] &=& {\sf D}^{a_0 b_0}[\omega]\cdot\chi^{b_0}[A] + 
\dalembert\theta^{a_0} = \chi^{a_0}[A] + \dalembert\theta^{a_0} 
\nonumber \\
\chi^{\bar{a}}[A^\omega] &=& {\sf D}^{\bar{a}\bar{b}}[\omega] \cdot 
\chi^{\bar{b}}[A]
\label{6b} ,
\end{eqnarray}
where 
\begin{equation}
{\sf D}^{ab}[\omega] = (-2)\,\tr\left(T^a \omega T^b \omega^{\dagger}\right).
\label{6x}
\end{equation}
is the adjoint representation of the Abelian gauge rotation $\omega$.
This quantity describes the transformation of a matter field in the adjoint
representation, i.e.~it is block-diagonal and unity in the neutral sector,
whereas it constitutes a rotational matrix in the charged subspace.  
From the explicit calculation presented in appendix \ref{app:1} we 
obtain the simple transformation law for the FP matrix (\ref{6a}):
\begin{equation}
{\sf M}[A^\omega] = {\sf D}[\omega]\cdot{\sf M}[A]\cdot{\sf D}^{T}[\omega]
\label{6c}
\end{equation} 
where ${\sf D}^T$ denotes the transpose of ${\sf D}$. In particular, this
equation implies the invariance of $\det {\sf M}[A]$ under Abelian gauge
rotations, since $\det{\sf D} = 1$.

Let us also note that the Abelian and coset gauge fixing conditions become
completely independent once we implement the latter \emph{exactly} (see also
ref.~\cite{R1}). This can be achieved e.g.~by replacing the coset gf.~term in
the quantum action $S_q$ by a Fourier representation of the FP
(gauge fixing) delta function,
\[
\exp\left(-\frac{1}{2\hbar\bar{\alpha}} \int d^4 x\,
  \chi^{\bar{b}}\chi^{\bar{b}}\right)\longrightarrow \int {\cal D}\ch{\phi}
\exp\left(-\frac{i}{2\hbar} \int d^4x\,\phi^{\bar{b}} \chi^{\bar{b}}\right).
\]
Here, the auxiliary multiplier field $\ch{\phi}$ has to transform like a
charged matter field (cf.~(\ref{4})) under the residual symmetry
in order to keep $S_q$ and the path integral measure invariant. 
We may then evaluate the FP determinant on the gf.~manifold, where the 
factorisation into charged and neutral parts holds:
\begin{equation}
\det {\sf M}_{\rm FP} = \det (-\dalembert)\cdot\det {\sf M}_{\rm FP}^{\rm ch}.
\label{7}
\end{equation}

Returning to eq.~(\ref{6}) for the generating functional, we have shown that
the quantum action $S_q$ is always invariant under the Abelian transformation 
(\ref{3}). 
This entails that the residual symmetry is only broken
by the Abelian gauge fixing and the source terms. Exploiting this fact and the
invariance of the path integral measure in (\ref{6}), we can derive Ward
identities in the usual way: We consider an infinitesimal change (\ref{3}) of 
variables in the functional integral (which does not affect the value of $Z$) 
and set the variation of the Abelian gf.~and source terms to zero.\footnote{As 
usual, we discard the surface term in the variation of the Abelian source 
term.}  Replacing fields by derivatives w.r.t.~sources (when acting on the 
generating functional), we obtain eventually
\begin{equation}
\left\{\frac{1}{\hbar \alpha} \partial_\mu \dalembert \frac{\delta}{\delta 
j_\mu^{b_0}(x)} + \partial_\mu j_\mu^{b_0}(x) + g\cdot f^{\bar{a}b_0 \bar{c}}\, 
j_\mu^{\bar{a}}(x)\, \frac{\delta}{\delta j_\mu^{\bar{c}}(x)}\right\}
Z\left[\neu{j},\ch{j}\right] = 0.
\label{8} 
\end{equation} 
This result can be transformed into an identity for the effective action,
i.e.~the generating functional of 1PI Green's functions. Our conventions are
as follows: We define the generating functional $W[j]$ of \emph{connected}
Green's functions by $Z[j] = \exp(-\hbar^{-1} W[j])$.  The usual Legendre
transformation leads then to the classical field ${\sf A}$ and the effective
action $\Gamma[{\sf A}]$, respectively:
\begin{eqnarray}
{\sf A}_\mu^a(x) [j] &=& \left\langle A_\mu^a(x)\right\rangle_j = 
\frac{\delta}{\delta j_\mu^a(x)} \ln Z[j] = - \frac{1}{\hbar} 
\frac{\delta W[j]}{\delta j_\mu^a(x)}
\nonumber \\
\Gamma[\sf A] &=& W[j[{\sf A}]] + \hbar \int d^4 x\,\,j_\mu^a[{\sf A}]\, 
{\sf A}_\mu^a.
\label{9}
\end{eqnarray}
With this convention, the Ward identity for $\Gamma$ becomes
\begin{equation}
\frac{1}{\alpha} \dalembert \partial_\mu {\sf A}_\mu^{b_0}(x) + 
\partial_\mu \frac{\delta \Gamma}{\delta {\sf A}_\mu^{b_0}(x)} + 
g\cdot f^{\bar{a} b_0 \bar{c}} \frac{\delta \Gamma}{\delta 
{\sf A}_\mu^{\bar{a}}(x)}{\sf A}_\mu^{\bar{c}}(x) = 0.
\label{10}
\end{equation}

Let us note that we can easily extend this identity to include ghost 
fields $(\eta,\bar{\eta})$, if we prefer to represent the FP determinant in 
$S_q$ by a ghost integral in the usual way:
\begin{equation}
\exp\left(\Tr\ln{\sf M}[A]\right) = \int \mathcal{D}(\eta,\bar{\eta})
\exp\left(-\frac{1}{\hbar}\int \bar{\eta}^a \cdot{\sf M}^{ab}[A]\cdot\eta^{b}
\right).
\label{10a}
\end{equation} 
From the transformation property (\ref{6c}) of the FP matrix, we infer that
the ghosts have to be rotated as charged and neutral (scalar) matter fields
w.r.t.~the residual Abelian symmetry, i.e.
\begin{eqnarray}
\eta^a \rightarrow {\sf D}^{ab}[\omega]\cdot \eta^b \qquad &;& \qquad
\eta = \eta^a T^a \rightarrow \omega\cdot\eta\cdot\omega^{\dagger} \nonumber \\
\bar{\eta}^a \rightarrow {\sf D}^{ab}[\omega]\cdot \bar{\eta}^b \qquad &;& 
\qquad \bar{\eta} = \bar{\eta}^a T^a \rightarrow \omega\cdot\bar{\eta}\cdot
\omega^{\dagger} 
\label{10b}.
\end{eqnarray}
In this way, the path integral measure and the quantum action $S_q$ 
(with the FP determinant replaced by the ghost term) will remain invariant
under Abelian transformations. In addition, however, we will have to 
couple sources to the ghosts in order to include them in the effective action.
The variation of these source terms under the gauge rotation
(\ref{10b}) contributes to the Ward identity,
\begin{eqnarray}
\frac{1}{\alpha} \dalembert \partial_\mu {\sf A}_\mu^{b_0}(x) + 
\partial_\mu \frac{\delta \Gamma}{\delta {\sf A}_\mu^{b_0}(x)} &+& 
g\cdot f^{\bar{a} b_0 \bar{c}}\Bigg( \frac{\delta \Gamma}{\delta 
{\sf A}_\mu^{\bar{a}}(x)}{\sf A}_\mu^{\bar{c}}(x) + \nonumber \\
{} && {} + \frac{\delta\Gamma}
{\delta_r \eta^{\bar{a}}(x)}\,\eta^{\bar{c}}(x) - \bar{\eta}^{\bar{a}}(x)\,
\frac{\delta\Gamma}{\delta_l \bar{\eta}^{\bar{c}}(x)} \Bigg)
 = 0.
\label{10c}
\end{eqnarray}
The results (\ref{10},\ref{10c}) are a direct consequence of the invariance
of the quantum action (\ref{6}) under Abelian gauge transformations and thus
hold for \emph{all} Abelian gauges.


\section{Renormalisation}
\label{sec:3}

Let us now consider the consequence of the Ward identities (\ref{10}),
(\ref{10c}) for the renormalisation of our theory. 
The quantum corrections to the effective action, given by the difference 
between $\Gamma[{\sf A}]$ (\ref{9}) and the tree action (first and second 
term in the exponent of eq.~(\ref{6})), can be expanded in powers of the 
fields\footnote{For simplicity, we
do not introduce ghost fields and use a shorthand notation where
summation integration over all relevant indices is understood.}
\begin{equation}
\Delta\Gamma = \frac{1}{2}  \neu{\sf A}\cdot\neu{\Pi}\cdot
\neu{\sf A} + \frac{1}{2} \ch{\sf A}\cdot\ch{\Pi}\cdot
\ch{\sf A} + \frac{1}{2} g \mu^{\epsilon/2} 
\delta G\cdot \ch{\sf A}\ch{\sf A}\neu{\sf A} + \cdots
\label{11}
\end{equation}
We will assume that the vacuum polarisations $\Pi$ and the vertex correction
$\delta G$ have been calculated to a given order in a loop expansion and
divergent contributions are regularised in a gauge invariant way. The precise
regularisation prescription is not important for the following, but in order
to be specific in eq.~(\ref{11}), we use dimensional regularisation to
$d=4-\epsilon$ Euclidean spacetime dimensions. Note that in this case, the
cutoff $\epsilon\to 0$ is dimensionless, but an additional scale $\mu$ must be
introduced to keep the regularised coupling constant dimensionless in $d\neq
4$.

The counterterms are constructed from the divergent part of the loop
correction,
\begin{eqnarray}
- \left. \Pi_{\mu\nu}^{\rm (n,ch)}(p,\mu,\epsilon)\right|_{\rm div, trans}
&=& \delta Z^{\rm (n,ch)}(p,\mu,\epsilon) \cdot \left[D_{0,{\rm (n,ch)}}^{-1}
\right]_{\mu\nu}^{\rm trans}(p) \nonumber \\
-\left.\Pi_{\mu\nu}^{\rm (n,ch)}(p,\mu,\epsilon)\right|_{\rm div, long} 
&=& \delta Z_{\rm (\alpha,\bar{\alpha})}(p,\mu,\epsilon)\cdot
\left[D_{0,{\rm (n,ch)}}^{-1}\right]_{\mu\nu}^{\rm long}(p) \label{12} \\
- \left. \delta G_{\mu\nu\rho}(p,\mu,\epsilon)\right|_{\rm div} &=& 
\delta Z_g(p,\mu,\epsilon) \cdot \left[G_0\right]_{\mu\nu\rho}(p).
\nonumber
\end{eqnarray}
Here, $D_0^{-1}$ and $G_0$ denote the (inverse) gluon propagator and triple
gluon vertex, respectively, as obtained from expanding the initial action
(\ref{6}) in powers of the fields.\footnote{The precise
  definition of $D_0^{-1}$ and $G_0$ can be read off from eq.~(\ref{17}) below,
  with $\Gamma_B$ replaced by the initial quantum action (\ref{6}).}
Furthermore, some renormalisation prescription to fix the finite parts in the
factors $\delta Z$ is understood. Adding the counterterms to the initial
action in (\ref{6}) yields the same action in terms of the \emph{bare} fields
given by ($Z_i = 1 + \delta Z_i$)
\begin{equation}
\begin{array}{l@{\;=\;}l@{\quad ; \quad}l@{\;=\;}l}
A_B^{\rm (ch)} & Z_{\rm ch}^{1/2}\cdot\ch{A} &
A_B^{\rm (n)} & Z_{\rm n}^{1/2}\cdot\neu{A} \\
\alpha_B & Z_\alpha Z_{\rm n}\cdot \alpha &
\bar{\alpha}_B & Z_{\bar{\alpha}} Z_{\rm ch}\cdot \bar{\alpha} \\
g_B & \multicolumn{3}{l}{g \mu^{\epsilon/2}\cdot Z_g Z_{\rm ch}^{-1} 
Z_{\rm n}^{-1/2}}  
\end{array}.
\label{13}
\end{equation}
This bare action must now be used as starting point to calculate the loop
corrections, and by construction, it will yield \emph{finite renormalised}
Green's functions\footnote{up to the loop order from which the counterterms
  were calculated.} when the bare quantities are re-expressed through the
renormalised fields and $Z$-factors.

Let us denote by $\Gamma_B[A_B,g_B,\alpha_B]$ the effective action obtained in
a loop expansion from the bare initial action.
It is understood that $\Gamma_B$ is rendered finite by the same gauge
invariant regularisation prescription that was already used for the extraction
of the counterterms. Note that this also introduces a scale $\mu_B$ for the
bare coupling $g_B$, which is however related to the scale $\mu$ in the
counterterms by the renormalisation prescription. If we want e.g.~the loop
divergences in $\Gamma_B$ to be just cancelled (i.e.~without finite parts) by
the counterterms encoded in the $Z's$, we must have $\mu_B = \mu$.

Since the regularisation does not spoil gauge symmetry, we can derive a Ward
identity on $\Gamma_B$ exactly as in (\ref{10}). Replacing finally bare
quantities by renormalised ones with the help of proper $Z$-factors, we
obtain:
\begin{equation}
\frac{1}{Z_\alpha \alpha}\cdot\dalembert \partial_\mu {\sf A}_\mu^{b_0}(x)
+ \partial_\mu \frac{\delta \Gamma_B}{\delta {\sf A}_\mu^{b_0}(x)} + 
\frac{Z_g}{Z_{\rm ch}}\cdot g \mu^{\epsilon/2}\,f^{\bar{a}b_0\bar{c}}\, 
\frac{\delta \Gamma_B}{\delta {\sf A}_\mu^{\bar{a}}(x)} 
{\sf A}_\mu^{\bar{b}}(x) = 0.
\label{14}
\end{equation}  
Here, we have not distinguished between $\mu$ and $\mu_B$ since any deviation
could be absorbed in the renormalisation prescription. It should be stressed
once again that the bare effective action $\Gamma_B$ already contains the
counterterms if re-expressed in terms of the renormalised quantities and
$Z$-factors. Thus, differentiating $\Gamma_B$ w.r.t.~the renormalised fields
yields finite, renormalised Green's functions when the cutoff is removed
($\epsilon\to 0$). As a consequence,
\begin{equation}
\lim\limits_{\epsilon\to 0}Z_\alpha(p,\mu,\epsilon) = \mbox{finite} 
\qquad ; \qquad
\lim\limits_{\epsilon\to 0}\frac{Z_g(p,\mu,\epsilon)}{Z_{\rm ch}(p,\mu,
\epsilon)} = \mbox{finite}
\label{15}
\end{equation}
since all other quantities in (\ref{14}) are finite as $\epsilon\to 0$.  In
the \emph{minimal subtraction scheme} (and in practice also in all other
commonly used regularisation prescriptions), this entails
\begin{equation}
Z_\alpha(p,\mu,\epsilon) = 1 
\qquad ; \qquad
\frac{Z_g(p,\mu,\epsilon)}{Z_{\rm ch}(p,\mu,\epsilon)} = 1.
\label{16}
\end{equation}
These considerations can be straightforwardly extended to the case where
FP ghosts are introduced. The following additional relations 
between bare and renormalised quantities arise
\begin{equation}
\begin{array}{l@{\;=\;}l@{\quad ; \quad}l@{\;=\;}l}
\eta_B^{\rm (ch)} & \tilde{Z}_{\rm ch}^{1/2}\cdot\ch{\eta} &
\eta_B^{\rm (n)} & \tilde{Z}_{\rm n}^{1/2}\cdot\neu{\eta} \\
\bar{\eta}_B^{\rm (ch)} & \tilde{Z}_{\rm ch}^{1/2}\cdot\ch{\bar{\eta}} &
\bar{\eta}_B^{\rm (n)} & \tilde{Z}_{\rm n}^{1/2}\cdot\neu{\bar{\eta}} \\
g_B & \multicolumn{3}{l}{g \mu^{\epsilon/2}\cdot \tilde{Z}_g 
\tilde{Z}_{\rm ch}^{-1} Z_{\rm n}^{-1/2}}  
\end{array}
\label{16a}
\end{equation}
and from the universality of the renormalised coupling constant (as defined
from the triple gluon or photon-ghost vertex), we infer
\begin{equation}
\frac{\tilde{Z}_g(p,\mu,\epsilon)}{\tilde{Z}_{\rm ch}(p,\mu,\epsilon)} = 
\frac{Z_g(p,\mu,\epsilon)}{Z_{\rm ch}(p,\mu,\epsilon)} = 1.
\label{16b}
\end{equation}
In fact, this relation also follows from the above considerations applied
to the renormalised Ward identity including ghosts (cf.~eq.~(\ref{10c})), 
\begin{eqnarray}
\frac{1}{Z_\alpha \alpha}\cdot\dalembert \partial_\mu {\sf A}_\mu^{b_0}(x)
&+& \partial_\mu \frac{\delta \Gamma_B}{\delta {\sf A}_\mu^{b_0}(x)} + 
\frac{Z_g}{Z_{\rm ch}}\cdot g \mu^{\epsilon/2}\,f^{\bar{a}b_0\bar{c}}\, 
\frac{\delta \Gamma_B}{\delta {\sf A}_\mu^{\bar{a}}(x)} 
{\sf A}_\mu^{\bar{b}}(x) + 
\label{16c} \\
{}&+& \frac{\tilde{Z}_g}{\tilde{Z}_{\rm ch}} \cdot g \mu^{\epsilon/2}\,
f^{\bar{a}b_0\bar{c}}\Bigg(\frac{\delta \Gamma_B}{\delta_r\eta^{\bar{a}}(x)}\,
\eta^{\bar{c}}(x) - \bar{\eta}^{\bar{a}}(x)\,\frac{\delta\Gamma_B}
{\delta_l\bar{\eta}^{\bar{c}}(x)}\Bigg) = 0.
\nonumber
\end{eqnarray}


\section{Abelian Dominance in Asymptotic Freedom}
\label{sec:4}

Let us briefly discuss the physical meaning of the Abelian Ward identities and
the non-renormalisation condition (\ref{16},\ref{16b}). As mentioned above, 
the derivatives of the bare regularised action w.r.t.~the regularised fields 
are finite when the cutoff is removed. Transforming to momentum space, this
entails that
\begin{eqnarray}
\left.\frac{\delta^2 \Gamma_B}{\delta {\sf A}_\mu^a(x)
\delta {\sf A}_\nu^b(y)}\right|_0 &=& \int \frac{d^d p}{(2 \pi)^d}\, 
e^{-i p (x-y)} \left[D^{-1}\right]_{\mu\nu}^{ab}(p) 
\equiv \hbar \left\langle A_\mu^a(x) A_\nu^b(y)\right\rangle^{-1}_{\rm ren}
\nonumber \\
\left.\frac{\delta^3\Gamma_B}{\delta {\sf A}_\mu^{a_0}(x) 
\delta {\sf A}_\nu^{\bar{b}}(y)\delta{\sf A}_\rho^{\bar{c}}(z)}\right|_0 &=& 
g \mu^{\epsilon/2} \int \frac{d^d(p,q)}{(2\pi)^{2d}}\,e^{-i p (x-y) 
-i q (x-z)}\cdot i\,G^{a_0\bar{b}\bar{c}}_{\mu\nu\rho}(p,q)
\label{17}
\end{eqnarray}
where the kernels $D^{-1}$ and $G$ depend on $p,\mu$ and the regulator
$\epsilon$, but are \emph{finite} when the cutoff is removed ($\epsilon\to
0$).  Note that $G$ is (up to a factor $-\hbar$) the momentum space
representation for the 1PI triple gluon vertex with external lines removed,
while $D$ is simply the gluon two-point function, as indicated above.

Ward identities for $\neu{D}$ and $G$ can be easily derived from (\ref{14}) by
differentiating w.r.t.~${\sf A}$ and then setting all fields to zero. We find
the colour structure
\begin{equation}
\left[\neu{D}\right]^{\bar{a}\bar{b}}_{\mu\nu}(p) = 
\delta^{\bar{a}\bar{b}}\cdot D_{\mu\nu}(p) \qquad ; \qquad 
G^{a_0\bar{b}\bar{c}}_{\mu\nu\rho}(p,q) = 
f^{a_0\bar{b}\bar{c}}\cdot G_{\mu\nu\rho}(p,q)
\label{18}
\end{equation}
to all orders in perturbation theory. Futhermore, the Ward identities
constrain the longitudinal parts of the Green's functions. For the photon
propagator, we have
\begin{equation}
p_\mu \left[\neu{D}\right]^{-1}_{\mu\nu}(p,\mu,\epsilon) - \frac{p^2 p_\nu}
{Z_\alpha(p,\mu,\epsilon) \alpha} = 0.
\label{19}
\end{equation} 
Since all other quantities in this equation are finite as $\epsilon\to 0$, we
find $Z_\alpha \equiv 1$ in the MS-scheme, which means that the neutral gauge
fixing parameter is not renormalised in Abelian gauges. Note also that the
longitudinal part of the solution to (\ref{19}),
\begin{equation}
\neu{D_{\mu\nu}}(p,\mu,\epsilon\to 0) = \left(\delta_{\mu\nu} - 
\frac{p_\mu p_\nu}{p^2}\right) \neu{D}(p^2,\mu) + 
\alpha\,\frac{p_\mu p_\nu}{p^4} 
\label{20}
\end{equation}
is exhausted by the tree-level propagator, so that there are no quantum
corrections to the longitudinal part of the two-point function. In particular,
the photon vacuum polarisation must always be
\emph{transversal}.\footnote{This is of course a consequence of the Lorentz
  gauge in the Abelian sector, but it holds independently of the coset
  gf.~condition, i.e.~also for non-linear gauges like the maximal Abelian
  gauge.}

Turning to the vertex Ward identity, we obtain
\begin{equation}
(p+q)_\mu\,G_{\mu\nu\rho}(p,q) + \frac{Z_g}{Z_{\rm ch}} \left( \left[\ch{D}
\right]^{-1}_{\nu\rho}(q) - \left[\ch{D}\right]^{-1}_{\rho\nu}(p)\right) = 0.
\label{21}
\end{equation}
Once again, this implies $Z_g = Z_{\rm ch}$ when the cutoff is removed (in the
MS-scheme, this holds for all values of the regulator $\epsilon$).  Notice
that this can be interpreted as a form of \emph{Abelian dominance} for
perturbation theory and asymptotic freedom: The relation between bare and
renormalised coupling in Abelian gauges is simply
\begin{equation}
g_B  = g \mu^{\epsilon/2}\cdot Z_{\rm n}^{-1/2}(p,\mu,\epsilon)
\label{22}
\end{equation}
where $Z_{\rm n}$ is determined by the Abelian vacuum polarisation alone. This
should not be taken to mean that charged fields can be neglected altogether,
but we only need to consider diagrams with (two) neutral \emph{external}
photon lines in order to obtain the full $\beta$-function. Stated differently,
we can extract asymptotic freedom from the correlator $\langle\neu{A}\neu{A}
\rangle$
alone, where only the diagonal part of the field configuration enters, but
these configurations are of course sampled with the full non-Abelian weight.
This is similar to the phenomenon of \emph{Abelian dominance} observed on the
lattice for certain low-energy observables like the Wilson loop in the 
maximal Abelian gauge.

Let us finally mention that similar identities can be derived in the case 
that explicit ghost fields are introduced and the coupling constant is 
defined by the ghost-photon vertex.
We may define scalar ghost propagators $D_{\rm gh}(p)$ and the ghost-photon
vertex function $G_\mu(p,q)$ exactly as in (\ref{17}) with the only 
difference that there is no Lorentz index for the scalar ghost fields. 
The corresponding Ward identity for the vertex is easily derived from
(\ref{16c}), i.e.
\begin{equation}
(p+q)_\mu\,G_\mu(p,q) + \frac{\tilde{Z}_g}{\tilde{Z}_{\rm ch}}
\left[D_{\rm gh}^{-1}(p) - D_{\rm gh}^{-1}(q)\right] = 0,
\label{23}
\end{equation}
which again yields the cancellation $\tilde{Z}_g(p,\mu,\epsilon) = 
\tilde{Z}_{\rm ch}(p,\mu,\epsilon)$. 
In ref.~\cite{R1}, an explicit calculation in the 
maximal Abelian gauge confirmed the validity of the Ward identity (\ref{23}) 
and the cancellation $\tilde{Z}_g = \tilde{Z}_{\rm ch}$ to one loop order:
\begin{equation}
\tilde{Z}_{\rm ch}(p,\mu,\epsilon) = \tilde{Z}_g(p,\mu,\epsilon) = 
1 + \hbar g^2\, \frac{3-\alpha}{8 \pi^2\epsilon} + \mathcal{O}(\hbar^2)
\label{24}.
\end{equation}
Note that the counterterms (and thus the $Z$-factors) are independent of $p$
and $\mu$ in the MS scheme. Furthermore, the effective action of \cite{R1}
yields a transversal photon vacuum polarisation with
\begin{equation}
Z_{\rm n}(p,\mu,\epsilon) = 1 + \hbar g^2 \frac{11 \kappa}{24 \pi^2\epsilon}
+ \mathcal{O}(\hbar^2) \qquad ; \qquad \kappa = C_2(SU(2)) = 2
\label{25}.
\end{equation}
From (\ref{22}) this gives indeed the correct one-loop $\beta$-function for 
$G=SU(2)$ from Abelian correlators alone.


\section{ Conclusions } 
\label{sec:concl}

Abelian gauges generally help to reduce Yang-Mills theory to a (non-standard) 
Abelian model of photons and charged matter. In the present work we have 
studied this reduction on a quantum level. While the Cartan decomposition 
initially complicates the theory (there are more renormalisation constants), 
this is compensated by gaining simple symmetry relations in the Abelian sector.
As a consequence, Abelian gauges are very useful whenever the physics under 
consideration can essentially be described by the effective Abelian theory. 

On the lattice, such an \emph{Abelian dominance} has been observed 
approximately in the low energy regime, but only for special gauges such as 
the maximal Abelian gauge. In the present work we have investigated the use of 
Abelian gauges in the asymptotic freedom regime. We find that Ward identities
associated with the residual Abelian symmetry put stringent restrictions on 
some of the Green's functions, leading to powerful relations between the 
renormalisation constants.
More precisely, we have proved that in any Abelian gauge, the correct 
perturbative $\beta$-function of Yang-Mills theory can be obtained from the 
Abelian gluon propagator alone. The effects of the non-Abelian gauge fields 
are entirely absorbed in the arising effective Abelian gluon propagator. 
This result can be interpreted as Abelian dominance in the high-energy regime. 
We are thus led to the conclusion that Abelian gauges can be convenient not 
only in the confinement region, but also in the asymptotic freedom regime.


\subsection*{Acknowledgment}
We would like to thank R.~Alkofer and M.~Engelhardt for carefully reading the
manuscript and helpful discussions.


\begin{appendix}

\section{Transformation Property of the FP Matrix}
\label{app:1}

We start from the definition (\ref{6a}) of the FP matrix with a gauge rotated
argument, 
\begin{equation}
{\sf M}^{ab}[A^\omega] = \left.\frac{\delta \chi^a[(A^\omega)^\Omega]}
{\delta \varphi^b}\right|_{\varphi=0} =
\left.\frac{\delta \chi^a[(A^{\tilde{\Omega}})^\omega]}{\delta \varphi^b}
\right|_{\varphi=0}  
\end{equation}
where $\omega = e^{- g \theta^{b_0} T^{b_0}}$ denotes an arbitrary Abelian 
gauge transformation and $\Omega = e^{- g \varphi^b T^b}$. In the second step,
we have introduced
\begin{equation} 
\tilde{\Omega} = \omega^{\dagger}\cdot \Omega\cdot\omega = 
e^{- g \tilde{\varphi}^c T^c}.
\label{app_0}
\end{equation}
This allows us to apply the special transformation properties (\ref{6b}) 
of Abelian gauges, $\chi^a[A^\omega] = {\sf D}^{ab}[\omega]\cdot\chi^b[A] + 
\mbox{const}[\omega]$, 
with the rotational matrix  ${\sf D}$ defined in (\ref{6x}). Upon 
applying the functional chain rule, we find 
\begin{equation}
{\sf M}^{ab}[A^\omega] = {\sf D}^{ac}[\omega] \cdot\left.\frac{\delta
\chi^c[A^{\tilde{\Omega}}]}{\delta \tilde{\varphi}^d}\right|_{\varphi=0}\cdot
\left.\frac{\delta\tilde{\varphi}^d}{\delta\varphi^b}\right|_{\varphi=0}.
\label{app_1}
\end{equation}
From the definition (\ref{app_0}), we observe that 
$\Omega=\unit \iff \tilde{\Omega}=\unit$, or $\varphi=0 \iff 
\tilde{\varphi} = 0$. The second factor in (\ref{app_1}) 
is then simply the FP matrix evaluated at the non-rotated argument $A_\mu$.
To complete the proof of the transformation property (\ref{6c}) quoted in 
the main text, we only have to show that
\begin{equation}
\left.\frac{\delta\tilde{\varphi}^a}{\delta\varphi^b}\right|_{\varphi=0}
= (-2)\,\tr(T^b\omega T^a\omega^{\dagger}) = {\sf D}^{ba}[\omega],
\end{equation} 
which follows, after a simple calculation, from the definition (\ref{app_0}).  
\end{appendix}



\end{document}